# Blockchain-Powered Software Defined Network-Enabled Networking Infrastructure for Cloud Management

Praveen Fernando
Department of Computer and Information Technology
Purdue University
West Lafayette, IN
ferna159@purdue.edu

Jin Wei
Department of Computer and Information Technology
Purdue University
West Lafayette, IN
kocsis0@purdue.edu

*Abstract— Cloud architecture has become a valuable solution for different applications, such as big data analytics, due to its high-degree of availability, scalability and strategic value. However, there still remain challenges in managing cloud architecture, in areas such as cloud security. In this paper, we exploit software-defined networking (SDN) and blockchain technologies to secure cloud management platforms from a networking perspective. We develop a blockchain-powered SDN-enabled networking infrastructure in which the integration between blockchain-based security and autonomy management layer and multi-controller SDN networking layer is defined to enhance the integrity of the control and management messages. Furthermore, our proposed networking infrastructure also enables the autonomous bandwidth provisioning to enhance the availability of cloud architecture. In the simulation section, we evaluate the performance of our proposed blockchain-powered SDN-enabled networking infrastructure by considering different scenarios.*

## I. Introduction

"Cloud" is a vast network of remote servers around the globe, which are interconnected and meant to operate as a single ecosystem. The market penetration for cloud services has been increasing in recent years due to its extensive benefits, such as high-valued flexibility, efficiency and strategic value. Due to its high penetration signified by the benefits, cloud architecture has become a "necessity" more than just a "choice" nowadays. However, there still remain essential challenges for managing the cloud architecture, in areas such as cloud security that can be compromised due to various factors. Cloud servers, which are divided geographically, typically use their dedicated networking infrastructure. For example, Google Cloud Platform uses their own Fiber to inter-connect various geographically separated facilities. When Cloud Service Providers (CSPs) can't sufficiently support their own cloud networking infrastructure, they may rely on a third party for networking services. This reliance on third-party networking services, whose trustworthiness cannot be guaranteed, can bring in additional security concerns. Various techniques have been proposed to enhance the security of the networking support of the cloud architecture. The authors in [1] proposed a mechanism to detect conflicting SDN related flow rules in a highly dynamic cloud computing environment. In [2] Chowhary *et al.* proposed a game-theoretic approach to detect DDOS attacks in a cloud environment by leveraging the programmability of SDN. These techniques are effective in managing the cloud environment. However, since the critical state variables are stored within a controller, the sophisticated attacks can be launched to manipulate these variables. An SDN-based framework was proposed in [3] for cloud security, which utilizes a centralized Security Manager to detect the falsely injected packets. This framework is effective when the centralized security manager works normally. However, due to its centralized nature, the security manager is highly susceptible to single point of failure, which can result in compromising the entire security layer. To address these issues, in this paper we develop a networking infrastructure to improve the transparency of the networking entities and enhance the security of the networking service. Our proposed networking solution can be applied for both private and public clouds. Furthermore, our networking infrastructure is developed by exploiting SDN and blockchain technologies.

SDN is a promising technology that removes the control plane from individual switches and transfer it to a centralized controller which possesses the topological view of the entire network [4]. This technology outperforms the conventional decentralized TCP/IP networking infrastructure by enabling comprehensive end-to-end controls with low-operational cost, having programmable hardware, and achieving automated decision making in network management. The rich southbound APIs are used to realize the network as well as push necessary configurations to the data plane switches. Therefore, in recent decades, SDN technologies have gained widespread popularity in different application fields [5], in which SDNs are used to establish the connectivity and traffic shaping amongst geographically separated hosts (i.e. physical servers) as well as logically separated tenants (i.e. virtual machines) in cloud architecture. In this work, we design a multi-controller SDN networking layer with OpenFlow 1.3 [6] as the southbound protocol. For simulation purposes, we have used Mininet, which is a Rapid Prototyping framework for SDN, with Ryu Controller [7]. Additionally, Ofsoftswitch13 [8] is used as the OpenFlow switch for simulation.

Blockchain is an emerging technology, which can be considered as an immutable and decentralized digital ledger [9]. A blockchain is a growing list of records, called blocks, which are linked via cryptography. Each block mainly contains the hash value of the previous block, the transaction data, and the timestamp. Due to the inherent cryptographic chaining, if a malicious party tries to manipulate certain transaction data, it will causes the changes of the hash values of the block containing the transaction and those of all the subsequent blocks, which can be easily detected. Therefore, generally speaking, blockchain technology provides a very promising solution for integrity and security. Amongst various existing blockchain platforms, Ethereum and Bitcoin are two of the most widely adopted ones [12] [13]. Compared with Bitcoin, Ethereum platform provides scripting functionalities via smart contracts. In our proposed networking solution, we exploit Ethereum blockchain and impose security functionalities automatically via smart contracts. The authors would like to claim that the technologies presented in this paper have been include in a provisional patent [11].

The next section describes our proposed blockchain powered SDN-enabled networking infrastructure. Simulation and results are shown in Section III followed by final remarks and the conclusion in Section IV.

## II. OUR PROPOSED BLOCKCHAIN-POWERED SDN-BASED NETWORKING INFRASTRUCTURE

The overview of our proposed blockchain-powered SDN-enabled networking infrastructure is illustrated in Fig. 1. Our infrastructure mainly comprises two layers: (1) multi-controller SDN networking layer, and (2) blockchain-based security and autonomy management layer. The interaction between these two layers are realized by connecting SDN controllers via a blockchain-based autonomous security mechanism. Overall, our networking infrastructure is developed to realize three main functionalities: (1) integrity verification for control and management commands for cloud platforms, (2) identification of the malicious hosts abusing the cloud platform, and (3) enhance the availability of the cloud platform via autonomous bandwidth provisioning.

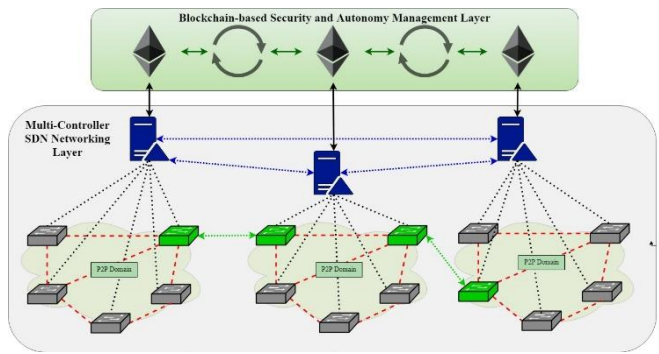

Fig. 1: Overview of our blockchain-powered SDN-enabled networking infrastructure.

### A. Multi-Controller SDN Networking Layer

Considering the scalability issue of the conventional centralized SDN, in our multi-controller SDN networking layer, OpenFlow-enabled switches span across different domains in each of which there is an SDN controller. To realize networking, both intra-domain and inter-domain communication are enabled as shown in Fig. 1. Different domains are connected with each other in a decentralized manner via domain-edge switches, which forms the data plane back-bone network, while different controllers are connected with each other in the vicinity to form the decentralized controller plane.

Furthermore, there are primarily two strategies for designing multi-controller SDN networking layer: (1) *horizontal architecture*, in which the controllers are interconnected in a peer-to-peer fashion, and (2) *hierarchical architecture,* in which all the controllers are connected to one centralized controller which is responsible for controller management. One major drawback of the hierarchical architecture is that the centralized controller is highly vulnerable to single point of failure. Therefore, we design our multi-controller SDN networking layer by adopting the horizontal architecture.

### B. Interaction with Blockchain-Based Security and Autonomy Management Layer

#### 1) Integrity Verification Mechanism

In our networking infrastructure, a verification mechanism is designed to verify the integrity of the critical control and management commands for the cloud platform. This mechanism is realized via the interaction between the multi-controller SDN networking layer and the blockchain-based security and autonomy management layer. As illustrated in Fig. 2, whenever an SDN controller is ready to send a control/management command, it generates a combined hash of the command data and the timestamp associated with the command. In our work, the combined hash value is calculated via MD5 message-digest algorithm [10]. Then the SDN controller records the hash value as a state variable in a smart contract in the blockchain-based layer. After generating and recording the hash value, the controller sends the control command with the timestamp to the targeted SDN controller (receiver). After receiving the command, the controller verifies the integrity of the command by hashing the received information, including the command and its associated timestamp, and checking against the blockchain smart contract to find out whether the aforementioned hash was previously generated in the system.

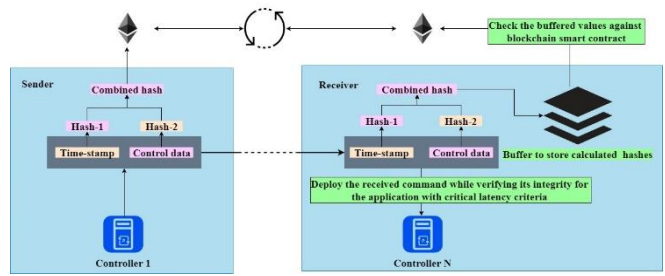

Fig. 2: Illustration of our integrity verification mechanism.

Additionally, the receiver can store the hash values in a buffer if it receives various control commands from the controllers. To avoid interrupting the communications with critical latency criteria, if it is necessary, the receiver can directly deploy received control commands without verifying the integrity. In the meantime, the buffered hash values are verified against the hash values recorded via the state variables in smart contracts.

*2) Detection Mechanism for Malicious Hosts*

To prevent malicious hosts from abusing the cloud platform, we develop a detection mechanism that detects malicious hosts by identifying the authorization of the individual hosts. As the first step for achieving the objective, currently we assume that every valid host acquires an IP address using Dynamic Host Configuration Protocol (DHCP) from the associated SDN controllers. Furthermore, we also assume that the Cloud Service Provider (CSP) has a knowledge of all the available MAC addresses in its management domain. Therefore, the CSP is able to bind an IP address to available MAC addresses and record the IP-MAC association in our blockchain-based security and autonomy layer, which ensures the immutable exchange of such information within controllers. A typical IP-MAC association table is illustrated in Table I.

Table I: Example of an IP-MAC association table.

| Host IP | MAC Address |
|---|---|
| 20.0.0.1 | 48-2C-6A-1E-59-3D |
| 20.0.0.2 | 48-2C-6A-1E-58-AC |
| 20.0.0.5 | 48-2C-6A-1E-A1-5D |

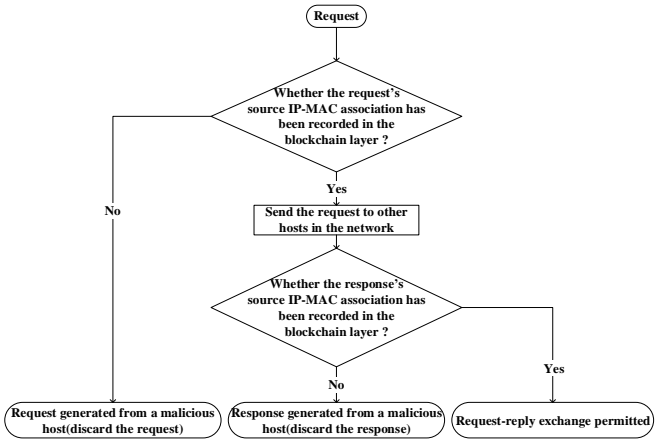

Fig. 3: Illustration of our detection mechanism for malicious hosts.

When a host requests an IP via DHCP, the SDN controller checks whether the host's MAC address is available in the IP-MAC association table in the blockchain-based layer. If it is, the controller concludes that the authorization of the host has been verified and release the corresponding IP address to the host. At the connection establishment phase, controller will also check the integrity of associated ARP messages which is illustrated in the fig. 3. By doing so, our networking solution ensures that two hosts cannot establish an end-to-end IP association without a qualifying ARP request-reply pair.

*3) Autonomous Bandwidth Provisioning Mechanism*

In our proposed networking solution, an autonomous bandwidth provisioning mechanism is developed to enhance the availability of the cloud platform. As illustrated in Fig. 4, in our proposed mechanism, the immutability and scripting functionality of the blockchain are exploited to impose Service Level Agreements (SLAs) into the networking logic. Additionally, the scripting functionality of the blockchain and the metering and queue functionality of the OpenFlow Switch Specification are employed to realize the real-time awareness of the available bandwidths of each links.

In this work, we assume that the malicious traffic flows are identified and discarded via our integrity verification mechanism and the detection mechanism for malicious hosts. Additionally, we consider the normal traffic flows can be classified into two categories: *guaranteed traffic flows* and *best-effort traffic flows*. This classification is automatically realized according to the flag segment in a SLA definition table. An example of SLA definition table is illustrated in Table II. In this table, the first row corresponds to guaranteed traffic that is identified by Flag 1, and the second row corresponds to best-effort traffic that is identified by Flag 0. When a provisioning request comes from a host, the Cloud Service Provider (CSP) imposes the associated SLA entry in the SLA definition table via the blockchain smart contract in our blockchain-based security and autonomy management layer. All the entries of the SLA definition table are recorded as *structs* and a mapping datatype is used to store all the entries in the blockchain-based layer with each data entry (of type *struct*) is mapped with an index (of type *unit256*). The entries of the SLA definition table are also updated in real-time via the blockchain smart contract in order to reflect the changes in underlying networking state.

Table II: Example of a SLA definition table.

| Source IP | Destination IP | SLA Bandwidth | Flag |
|---|---|---|---|
| a.a.a.a | b.b.b.b | 1 Mbps | 1 |
| c.c.c.c | d.d.d.d | 2 Mbps | 1 |
| e.e.e.e | f.f.f.f | 0 Mbps | 0 |

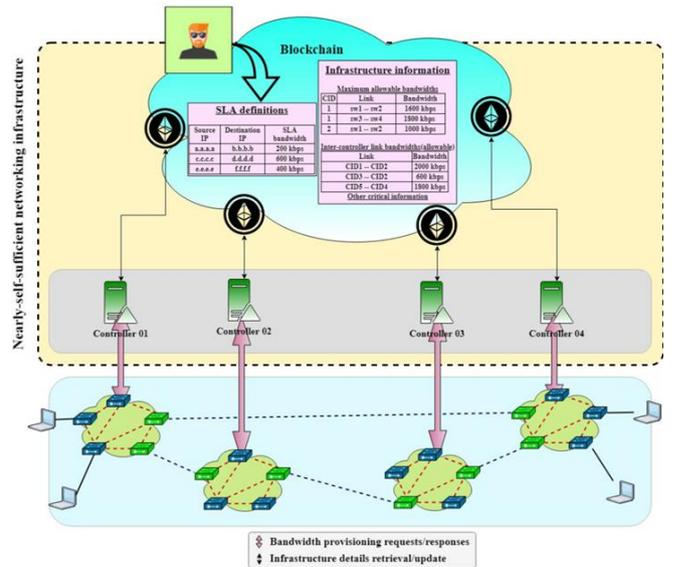

Fig. 4: Overflow of our autonomous bandwidth provisioning mechanism.

Then the host sends a provisioning request to establish the end-to-end connection and the SDN controllers deploy the requested flow according to the mechanism illustrated in Fig.

5. If it is a *best-effort* service request, the SDN controller routes the traffic along the shortest path from source to destination. To achieve this automatically, one critical data-structure, *traversal edge-switch matrix*, is defined via the blockchain smart contract. A certain entry of this matrix shows the corresponding edge-switch when traversing from one controller to another. This matrix is used to identify the two domain-edge switches corresponding to the domain of each controller that participates in the packet switching process for a certain source and destination. This information is fed to individual controller to find the local shortest paths of each controller.

If the request from the host is a guaranteed service request, the associated source SDN controller checks whether there is enough bandwidth available from source to destination for the guaranteed flow.

To automatically identify whether there is sufficient bandwidth available for the guaranteed traffic, two critical data-structures are defined via the blockchain smart contract: (1) *Inter-controller link bandwidth matrix* that records all the links connecting the SDN controllers and their available bandwidths, and (2) *Intra-controller link bandwidth matrix* that records the bandwidths of all the available links within the domain of individual controllers. These two matrices are used to track all the available bandwidths throughout the network and make decision on the bandwidth provisioning for a given pair of source and destination hosts. If there is sufficient bandwidth available, the provisioning request can be fulfilled successfully and the availability of the bandwidth is updated in the blockchain-based layer accordingly. If there is no sufficient bandwidth available, a decision making procedure is implemented on identifying an alternative *guaranteed* path. If there are no appropriate guaranteed path candidates available, traffic will be sent along the best-effort path. The data flow is automatically switched back to the *guaranteed* path when the bandwidth becomes sufficient. During this procedure, new provisioning requests are blocked to ensure that link bandwidths have been updated accurately before new provisioning decisions can be made.

### III. SIMULATION RESULTS

In this section, we evaluate the performance of our proposed blockchain-powered SDN-enabled networking infrastructure by considering two case studies. In Case A, we focus on demonstrating the performance of the verification mechanism in our networking infrastructure, which is designed to verify the integrity of the critical control and management commands for the cloud platform. In Case B, we focus on evaluating the performance of our proposed autonomous bandwidth provisioning mechanism.

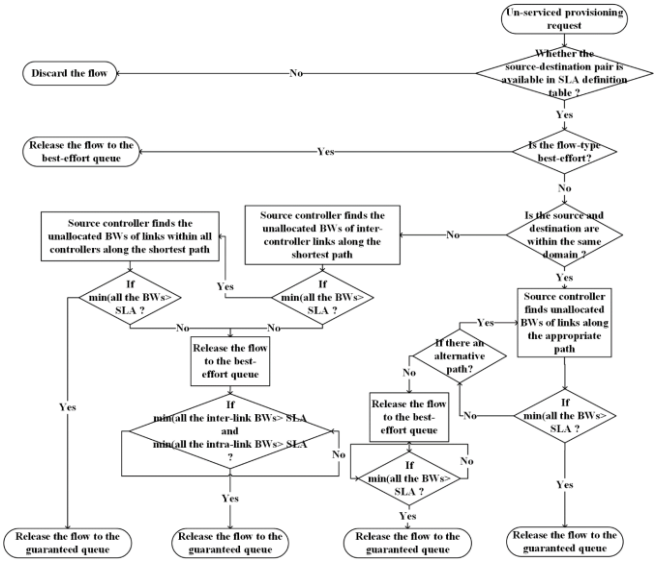

Fig. 5: Diagram for illustrating the mechanism of our autonomous bandwidth provisioning method.

In our proposed networking infrastructure, each port contains a queue to service the guaranteed traffic. There's no queue defined for the best-effort traffic. Figure 6 represents the bandwidth allocation inside a certain link. As shown in Fig. 6, the guaranteed traffic has a predefined maximum guaranteed bandwidth. Additionally, due to its nature of expansion and shrinking based on the throughput, the networking resources can be utilized efficiently without any resource exhaustion.

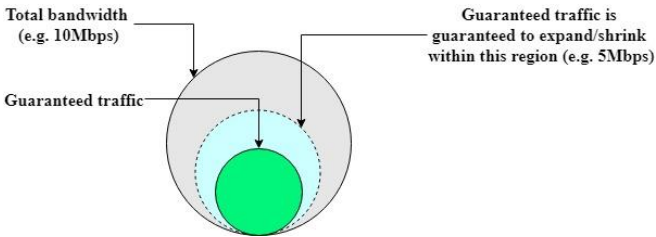

Fig. 6: Illustration of bandwidth allocation inside a link for a guaranteed traffic.

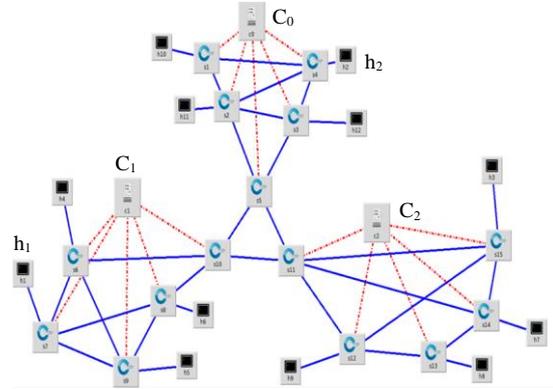

Fig. 7: Illustration of the topology of our multi-controller SDN-based networking layer in Case A.

In this case study, we consider that the multi-SDN networking layer comprises of three controllers, $C_0$, $C_1$, and $C_2$ as shown in Fig. 7. The three SDN domains associated with these individual controllers are connected with each other via domain-edge Switches $S_5$, $S_{10}$ and $S_{11}$ that are used for inter-domain traffic. The host ($h_1$) associated with $C_1$ intends to establish the communication with the host ($h_2$) associated with $C_0$. We assume that $h_1$ and $h_2$ do not have the knowledge about each other's MAC address. Initially, $h_1$ floods a control command, in this case an ARP request, requesting the MAC address of $h_2$ throughout all the controller domains. The ARP request is propagated to $C_0$ and $C_2$. The communication between the hosts associated with $C_0$ and $C_1$ is established after $h_2$ responds to the ARP request successfully. The screenshot of the terminals associated with the three SDN controllers are shown in Fig. 8, which presents the partial operations in our blockchain-based security and autonomy layer.

As shown in Fig. 8(a), When $C_1$ sends the ARP request of $h_1$ to other controllers, it also hashes the content and timestamp of the ARP request and records the combined hash value in the blockchain-based security and autonomy layer by uploading it as a state variable in the blockchain smart contract. From Fig. 8(b), we can see that, after receiving the ARP request, both the controllers $C_0$ and $C_2$ validate the integrity of the received command against the blockchain smart contract. After the cross-validation is completed successfully, the controller $C_0$ replies to the request as shown in Fig. 8(c). The integrity of response from $C_0$ is also verified at $C_1$ using the same validation mechanism.

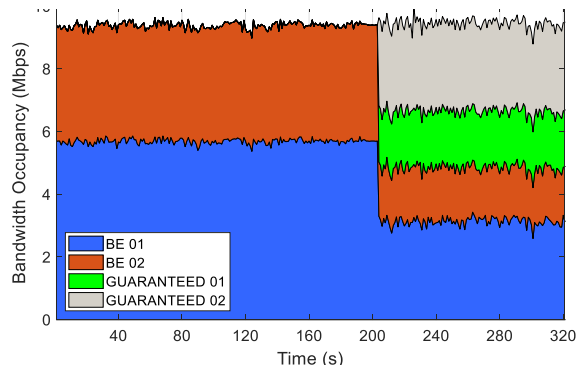

(a)

(b)

(c)

Fig. 8: Screenshot of the terminals for the three controllers: (a) the terminal for C1 when sending ARP request, (b) the terminals for C0 and C2 after receiving the ARP request, and (c) the terminal for C0 when ARP reply is generated and the terminal of C1 when ARP reply is verified upon reception.

*2) Case Study B*

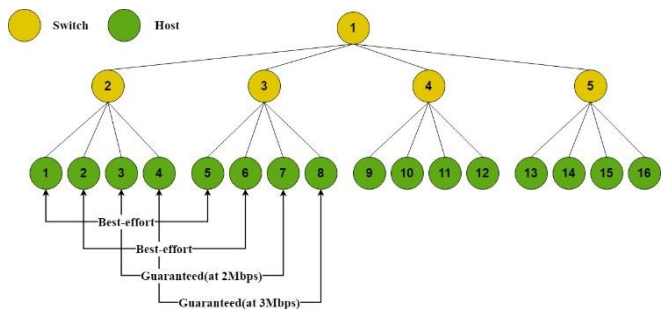

Fig. 9: Illustration of the topology of the networking layer in Case B.

In this case study, we consider that the networking layer has depth = 2 and fan-out = 4, which is shown in Fig. 9. We assume that maximum link bandwidth is 9.4 Mbps. The queue functionality of the OpenFlow Switch Specification in the networking layer is applied to achieve the guaranteed bandwidth of 5 Mbps for each port, which enables logically defining a tunnel for the *guaranteed* traffic. Four different service-request traffics are considered in this simulation, which is detailed in Table III.

Table III: Details of the service-request traffics.

| Flow | SLA Bandwidth | Traffic Type | Identification |
| --- | --- | --- | --- |
| Between Host 1 and Host 5 | 5.7 Mbps | *Best-effort (BE)* service request | BE 1 |
| Between Host 2 and Host 6 | 3.7 Mbps | *BE* service request | BE 2 |
| Between Host 3 and Host 7 | 1.8 Mbps | *Guaranteed* service request | Guaranteed 1 |
| Between Host 4 and Host 8 | 2.8 Mbps | *Guaranteed* service request | Guaranteed 2 |

As shown in Table III, the traffics BE 1 and BE 2 are the *best-effort* service-request traffics and the traffics Guaranteed 1 and Guaranteed 2 are the *guaranteed* service-request traffics. We assume that these four traffic flows are selected for the analysis in such a way that all of them are routed via the same shortest path. During the simulation, there only exist the traffics BE 1 and BE 2 from time t = 0s to 203s. The traffics Guaranteed 1 and Guaranteed 2 appear beginning from time step t = 203s. The bandwidth occupancy is shown in Fig. 10.

Fig. 10: Accumulative bandwidth occupancies for the service-request traffics.

As shown in Fig. 10, the *best-effort* service-request traffics BE 1 and BE 2 are able to occupy the entire available link bandwidth until $t = 203s$. The *guaranteed* service-request traffics appear at $t = 203s$, whose guaranteed bandwidths are 1.8 Mbps and 2.8 Mbps, respectively. These two guaranteed traffics use almost all the guaranteed bandwidth that is 5 Mbps. The immutability and the scripting functionality of the blockchain technology in our blockchain-based layer are exploited to securely and automatically impose the SLAs to the network logic. For example, in this case study the blockchain-based layer ensures that the total bandwidth of the flows in the guaranteed queue should be less or equal to guaranteed queue maximum bandwidth. The metering functionality of the OpenFlow Switch Specification in the networking layer is exploited to bandlimit the traffics Guaranteed 1 and Guaranteed 2 to 2 Mbps and 3 Mbps inside the guaranteed queue, respectively, which is implemented to prevent any potential resource exhaustion.

The packet-loss rates for the individual traffics are shown in Fig. 11. From Fig. 11, we can observe during the time window [$1s, 203s$], the packet-loss rates for the traffics BE 1 and BE 2 are neglectable. This is reasonable because that their requested bandwidths are 5.7 Mbps and 3.7 Mbps, respectively. The total requested bandwidth, 9.4 Mbps, is approximately equal to the total available link bandwidth, and thus these two traffics do not interfere each other. Additionally, we can observe significant packet-loss rate for BE 1 and BE 2 beginning from time $t = 203\,s$. This is because that the two *guaranteed* traffics, Guaranteed 1 and Guaranteed 2, which appear beginning from t = 204s, take up to 4.6 -Mbps bandwidth. Because of this the bandwidth available for the two *best-effort* traffics is reduced from 9.4 Mbps to 5 Mbps. Furthermore, we can observe that the packet-loss rates of the two *guaranteed* traffics are very low throughout the simulation, which illustrate the effectiveness of the metering and queue functionalities of the OpenFlow Switch Specification in the networking layer and the immutability and scripting functionality provided by the blockchain-based security and autonomy layer.

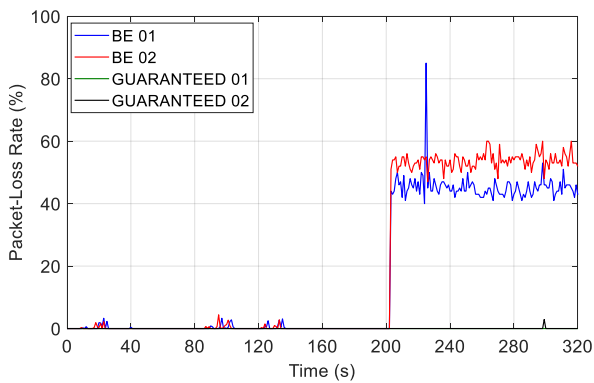

Fig. 11: Packet-loss rates for the individual service-request traffics.

## IV. CONCLUSIONS

In this paper, we have developed a blockchain-powered SDN-enabled networking infrastructure to enhance the security of the cloud management, such as integrity and availability. Our proposed networking infrastructure mainly consists of two layers: multi-controller SDN networking layer and blockchain-based security and autonomy layer. The integration of these two layers is designed to enhance the integrity of the control and management commands. Furthermore, our proposed networking infrastructure also enables the autonomous bandwidth provisioning to improve the availability of the cloud architecture. The simulation results in two case studies have illustrated the effectiveness of our proposed networking solution in enhancing the integrity and availability of the cloud management. In our ongoing work, we are developing a hardware-in-the-loop (HIL) testbed to further evaluate the performance of our work.

## V. ACKNOWLEDGEMENT

This research work was supported by NASA under Grant 80NSSC17K0530.